\documentclass[twocolumn,showpacs,preprintnumbers,amsmath,amssymb]{revtex4}

\usepackage{graphicx}
\usepackage{dcolumn}
\usepackage{bm}% bold math

\begin{document}

\title{Exactly solvable model of the 2D electrical double layer}

\author{L. {\v S}amaj}
\email{fyzimaes@savba.sk}
\affiliation{
Institute of Physics, Slovak Academy of Sciences\\
D\'ubravsk\'a cesta 9, 845 11 Bratislava, Slovak Republic
}

\author{Z. Bajnok}
\email{bajnok@elte.hu}
\affiliation{
Theoretical Physics Research Group, Hungarian Academy of Sciences\\
E\"otv\"os University, P\'azm\'any P\'eter s\'et\'any 1/A, 
1117 Budapest, Hungary
}

\date{\today}

\begin{abstract}
We consider equilibrium statistical mechanics of a simplified model
for the ideal conductor electrode in an interface contact with 
a classical semi-infinite electrolyte, modeled by the two-dimensional
Coulomb gas of pointlike $\pm$ unit charges 
in the stability-against-collapse regime 
of reduced inverse temperatures $0\le \beta<2$.
If there is a potential difference between the bulk interior
of the electrolyte and the grounded interface, the electrolyte
region close to the interface (known as the electrical double
layer) carries some nonzero surface charge density.
The model is mappable onto an integrable semi-infinite sine-Gordon 
theory with Dirichlet boundary conditions.
The exact form-factor and boundary state information gained from the mapping 
provide asymptotic forms of the charge and number density profiles of
electrolyte particles at large distances from the interface. 
The result for the asymptotic behavior of the induced electric potential,
related to the charge density via the Poisson equation, confirms 
the validity of the concept of renormalized charge and the
corresponding saturation hypothesis.
It is documented on the non-perturbative result for the asymptotic 
density profile at a strictly nonzero $\beta$ that the Debye-H\"uckel
$\beta\to 0$ limit is a delicate issue. 

\end{abstract}


\maketitle

\renewcommand{\theequation}{1.\arabic{equation}}
\setcounter{equation}{0}

\section{Introduction}
Asymmetric classical Coulomb mixtures, such as high\-ly charged
colloidal or polyelectrolyte suspensions, have attracted much
attention in the last years due to the appearance of various anomalous 
phenomena; for a nice review, see Ref. \onlinecite{Levin}.
The contemporary theoretical treatment of equilibrium statistical mechanics 
of asymmetric Coulomb mixtures is based on the concept of renormalized charge.
This concept has been introduced within the Wigner-Seitz cell models 
to describe an effective interaction between highly-charged ``macro-ions'' 
as a result of their strong positional correlations with the oppositely
charged ``micro-ions'' \cite{Manning,Alexander,Lowen,Trizac1}.
The idea of renormalized charge is usually documented in the infinite
dilution limit on a simplified model of a unique charged colloidal 
``guest'' particle immersed in a symmetric weakly-coupled electrolyte 
\cite{Belloni,Diehl,Trizac2,Aubouy}.
Plausible arguments were given to conjecture that the induced electric 
potential far from the guest colloid exhibits the form predicted by 
the Debye-H\"uckel (DH) theory [sometimes called the linear Poisson-Boltzmann 
(PB) theory], with a renormalized-charge prefactor.
Within the framework of the {\em nonlinear} PB theory, the renormalized 
charge saturates at some finite value when the colloidal bare charge 
goes to infinity \cite{Trizac2,Aubouy}.
A more general phenomenon of the saturation of the induced electric 
potential at each point of the electrolyte region was studied 
in Ref. \onlinecite{Tellez}. 

The idea of renormalized charge was developed within the linear
and nonlinear versions of the PB theory, which are rigorously valid
in the limit of the infinite temperature \cite{Kennedy}.
In order to treat correctly the Coulomb system at a finite temperature,
one has to go beyond these mean-field approaches by incorporating
electrostatic correlations among electrolyte particles.
As concerns the large-distance asymptotic behavior of the electric 
potential induced by a guest charge, this should affect both the
renormalized-charge prefactor (provided that the concept of
renormalized charge remains valid) as well as the correlation 
length of electrolyte particles which is expected to govern 
the exponential decay of the electric potential.
In spite of the existence of many phenomenological approximations
based on heuristic extensions of the mean-field theories \cite{Levin},
there is no chance to solve exactly a three-dimensional (3D) 
Coulomb fluid at some finite temperature.

The situation is more optimistic in the case of 2D Coulomb fluids 
consisting of charged constituents with logarithmic pairwise interactions. 
These systems maintain many generic properties, like screening and 
the related sum rules \cite{Martin}, of ``real'' 3D Coulomb fluids.  
The 2D Coulomb gas of symmetric $\pm$ unit {\em point-like} charges 
is stable against the collapse of positive-negative pairs of charges 
at high enough temperatures, namely for $\beta < 2$ where $\beta$ 
is the (dimensionless) inverse temperature or the coupling constant.
The collapse starts to occur at $\beta=2$: interestingly, although 
the free energy and each of the species densities diverge, 
the truncated Ursell correlation functions are finite 
at this inverse temperature.
The collapse $\beta=2$ point is exactly solvable due to its equivalence
with the free-fermion point of the Thirring fermionic representation of
the 2D Coulomb gas \cite{Cornu1,Cornu2}.
The exact solution involves the bulk thermodynamics for an infinite
system and special cases of the surface thermodynamics for a semi-infinite 
Coulomb gas in contact with an impermeable dielectric wall; 
for an exhaustive review, see Ref. \onlinecite{Jancovici92}.

Besides the Thirring fermion representation, the 2D Coulomb gas is 
equivalent to the 2D Euclidean, or (1+1)-dimensional quantum, 
sine-Gordon field theory with a conformal normalization 
of the cos-field \cite{Minnhagen}.
Although the bulk 2D sine-Gordon model has been known to possess
the integrability property from the seventies \cite{Zamolodchikov79}, 
the explicit exact solution of its ground-state characteristics 
was derived only quite recently due to a progress in the method of
Thermodynamic Bethe Ansatz (TBA) \cite{Destri,Zamolodchikov95}.
Based on the equivalence with the sine-Gordon theory, the bulk 
thermodynamic properties (free energy, internal energy, specific heat,
etc.) of the 2D Coulomb gas have been obtained exactly in the whole 
stability region of point-like charges $\beta<2$ \cite{Samaj00}.
From a gnoseological point of view, this is the only exactly solvable
case of a continuous (i.e. not on a lattice) fluid in more 
than one dimension.
Later on, the form-factor approach \cite{Smirnov,Lukyanov} was applied
to calculate the large-distance asymptotic behavior of the charge
\cite{Samaj02a} and number density \cite{Samaj02b} pair correlation 
functions in the bulk of the infinite 2D plasma.
The integrability of the 2D sine-Gordon theory was shown to be
preserved also in the half-space geometry with specific types
of boundary conditions \cite{Ghoshal1}.
The surface thermodynamic properties (surface tension) of the 2D
Coulomb gas in contact with an ideal conductor wall \cite{Samaj01a} and
an ideal dielectric wall \cite{Samaj01b} were obtained in the
whole stability range of $\beta<2$ through the mappings onto 
the boundary sine-Gordon model with Dirichlet and Neumann boundary 
conditions, respectively, with the aid of the known TBA results 
\cite{Skorik,Bajnok02} for these integrable boundary field theories.   

The exact treatment of the 2D Coulomb gas was recently extended also
to inhomogeneous situations of the present interest when one guest 
arbitrarily-charged particle is immersed in the bulk of an electrolyte
modeled by the 2D Coulomb gas.
At the free-fermion (collapse) point $\beta=2$, the considered problem
is solvable in the Thirring format \cite{Samaj05a} even for the guest 
$Q$-charged particle being of colloidal type, i.e. possessing 
a hard core of radius $a$ which is impenetrable to the electrolyte 
$\pm$ unit charges.
Based on an explicit formula for the electric potential induced
by the charged colloid in the electrolyte region, the concept of the
renormalized charge \cite{Belloni,Diehl,Trizac2,Aubouy}
was shown to fail in this strong-coupling regime.
On the other hand, in the limit $Q\to\infty$, the anticipated
phenomenon of the electric potential saturation \cite{Tellez} 
was confirmed at this free-fermion point.  
The special case of the {\em point-like} $(a=0)$ guest $Q$-charge 
is solvable, within the framework of the sine-Gordon format, inside 
the whole stability interval of the electrolyte $\beta<2$ \cite{Samaj05b}.
The explicit results for the asymptotic behavior of the induced
electric potential confirm the adequacy of the concept of
renormalized charge in this weak-coupling regime.
The exact results are rigorously valid provided that 
$\beta \vert Q\vert < 2$, i.e. when the guest $Q$-charge does not
collapse with an opposite unit charge (counterion) from the
electrolyte.
The possibility of an analytic continuation of the results beyond
the stability border $\beta \vert Q\vert = 2$ was conjectured
\cite{Samaj05b}, however, the validity of this ``regularization
hypothesis'' was not rigorously proved.
The restricted rigorous validity of the exact results to the region
$\beta \vert Q\vert <2$ prevents one from studying the saturation
phenomena in the limit $Q\to\infty$.

In order to avoid the collapse of point-like guest charges with
electrolyte counterions, one has to search for another integrable 
2D model with the guest charges uniformly smeared over a line manifold.
The simplest system of this kind is the 2D half-space Coulomb gas 
in contact with a plain hard wall carrying a uniform ``line''
charge density.
Although this model is exactly solvable at the free-fermion $\beta=2$ point
\cite{Cornu2}, it can be easily shown that its sine-Gordon 
formulation does not belong to the family of the boundary sine-Gordon
theories integrable at arbitrary $\beta<2$ \cite{Ghoshal1}.
Another way how to introduce an interface charge density is to consider
a simplified model for an electrode in contact with a classical
electrolyte: the half-space Coulomb gas bounded by an ideal conductor wall, 
with a potential difference $\varphi$ between the bulk interior of 
the electrolyte and the grounded interface \cite{Jancovici86,Forrester}.
As soon as $\varphi\ne 0$, the region of the Coulomb fluid close
to the interface (known as the electrical double layer)
carries some nonzero surface charge density.
The 2D version of the model was mapped in Ref. \onlinecite{Samaj01a}
onto the integrable half-space sine-Gordon model with a specific 
$\varphi$-dependent Dirichlet boundary condition.

The present paper concentrates just on this integrable model of 
the 2D electrical double layer and has two main aims.
The first aim is rather technical: we present a method for obtaining 
the charge and number density profiles of electrolyte particles at 
asymptotically large distances from model's interface.
This task is equivalent to the calculation of one-point functions
of bulk fields in the boundary sine-Gordon model with Dirichlet
boundary conditions.
Although the one-point functions of certain specific boundary field 
theories have already been analyzed using the truncated conformal
space approach and the form-factor expansion \cite{Dorey,Bajnok05},  
some additional generalizations and technicalities have to be
developed for the present model.
Among others we document on the exact non-perturbative asymptotic
form of the {\em number density} profile at large distances from 
the interface that the DH $\beta\to 0$ limit is a delicate point
which has to be taken with cautiousness.
The second aim consists in pointing out physical consequences of
the obtained exact results.
The knowledge of the asymptotic behavior of the charge density at
large distances from the interface enables us to derive 
the asymptotic large-distance tendency of the induced electric
potential toward its bulk value $\varphi$.
In the whole stability range of the electrolyte coupling $\beta<2$,
the asymptotic form of the electric potential coincides, up to
a renormalized-charge prefactor and the plasma correlation length, 
with the one obtained in the DH limit.
This result supports the general validity of the concept of renormalized
charge in the weak-coupling regime of the electrolyte.
The saturation hypothesis of the induced electric potential is also
confirmed. 

The article is organized as follows.
In Sec. II, we introduce the notation and briefly summarize 
important aspects of the mapping of the infinite 2D Coulomb gas
onto the bulk (1+1)-dimensional sine-Gordon theory.
Sec. III deals with the 2D electrical double layer of interest and
its mapping onto the semi-infinite sine-Gordon theory with Dirichlet
boundary conditions.
We would like to emphasize that the presentation in Secs. II and III 
is sketchy; for a detailed study of some specific points,
we quote relevant references.
Sec. IV presents the mean-field theories for the 2D electrical
double layer: the DH $\beta\to 0$ limit with the leading 
$\beta$-correction, and the nonlinear PB theory.
The crucial Sec. V is devoted to the derivation of the asymptotic
charge and number density profiles for the 2D electrical double layer,
in the whole stability region of inverse temperatures $\beta<2$.
The Thirring free-fermion point $\beta=2$ is discussed in Sec. VI.
A brief recapitulation and some concluding remarks are given 
in Sec. VII.

\renewcommand{\theequation}{2.\arabic{equation}}
\setcounter{equation}{0}

\section{The 2D Coulomb gas in the bulk}

\subsection{Basic definitions}
We start with a brief description of the classical Coulomb gas
formulated in the infinite 2D space of points ${\bf r} \in \mathbb{R}^2$.
It is realized as the limit of a finite system with 
periodic boundary conditions.
The system consists of point-like particles $\{ i \}$ of charge
$\{ q_i = \pm 1 \}$ (the elementary charge $e$ is set for simplicity 
to unity) immersed in a homogeneous medium of dielectric constant 
$\epsilon = 1$.
The interaction energy $E$ of a set of particles $\{ q_i, {\bf r}_i \}$
is given by
\begin{equation} \label{2.1}
E(\{ q_i, {\bf r}_i \}) = \sum_{i<j} q_i q_j 
v(\vert {\bf r}_i-{\bf r}_j\vert ) ,
\end{equation}   
where the electrostatic potential $v$ is the solution of 
the 2D Poisson equation
\begin{equation} \label{2.2}
\Delta v({\bf r}) = - 2 \pi \delta({\bf r}) 
\end{equation}
subject to the boundary condition
$\nabla v({\bf r}) \to 0$ as $\vert {\bf r} \vert \to \infty$.
Explicitly, one has
\begin{equation} \label{2.3}
v({\bf r}) = - \ln \left( \frac{\vert {\bf r} \vert}{r_0} \right) ,
\qquad r \in \mathbb{R}^2 .
\end{equation}
The free length constant $r_0$, which fixes the zero point of 
the Coulomb potential, will be set for simplicity to unity.
The Fourier transform of the 2D Coulomb potential (\ref{2.3}) exhibits
the form $1/\vert {\bf k} \vert^2$ with the characteristic singularity
at ${\bf k}\to {\bf 0}$.
This maintains many generic properties (like screening and the related
sum rules \cite{Martin}) of 3D Coulomb fluids with the interaction
potential $v({\bf r}) = 1/\vert {\bf r}\vert$, ${\bf r}\in \mathbb{R}^3$.

The system is treated in thermodynamic equilibrium, via the grand
canonical ensemble characterized by the (dimensionless) inverse
temperature $\beta$ and the couple of particle fugacities
$z_+ = z_- = z$.
The grand partition function is defined by
\begin{subequations} \label{2.4}
\begin{equation} \label{2.4a}
\Xi = \sum_{N_+=0}^{\infty} \sum_{N_-=0}^{\infty}
\frac{z_+^{N_+}}{N_+!} \frac{z_-^{N_-}}{N_-!} Q(N_+,N_-) , 
\end{equation}
where
\begin{equation} \label{2.4b}
Q(N_+,N_-) = \int_{\mathbb{R}^2} \prod_{i=1}^N {\rm d}^2 r_i
\exp \left[ - \beta E(\{ q_i,{\bf r}_i \}) \right]
\end{equation}
\end{subequations}
is the configuration integral of $N_+$ positive and $N_-$ negative
charges, and $N = N_+ + N_-$. 
For the considered case of point-like particles, the singularity of 
the Coulomb potential (\ref{2.3}) at the origin ${\bf r}={\bf 0}$ can 
cause the thermodynamic collapse of positive-negative pairs of charges. 
The stability regime against this collapse is associated with 
the 2D spatial integrability of the corresponding Boltzmann factor
$\exp[\beta v({\bf r})] = \vert {\bf r} \vert^{-\beta}$ at short distances,
and therefore corresponds to small enough inverse temperatures $\beta < 2$.
At large distances, the Coulomb interaction is screened to a
short-distance effective interaction of Yukawa type.  
The infinite system is homogeneous and translationally invariant. 
Denoting by $\langle \cdots \rangle_{\beta}$ the thermal average,
the number density of particles of one charge sign $q$ $(= \pm 1)$ is
defined by $n_q = \langle \sum_i \delta_{q,q_i} \delta({\bf r}-{\bf r}_i) 
\rangle_{\beta}$.
Due to the charge symmetry, $n_+ = n_- = n/2$ where $n$ is the total
number density of particles.
At the two-particle level, one introduces the two-body densities
$n_{qq'}(\vert {\bf r}-{\bf r}'\vert) = \langle \sum_{i\ne j}
\delta_{q,q_i} \delta({\bf r}-{\bf r}_i)  
\delta_{q',q_j} \delta({\bf r}'-{\bf r}_j) \rangle_{\beta}$, etc. 

\subsection{The sine-Gordon representation}
The infinite 2D Coulomb gas is mappable onto the bulk sine-Gordon
theory \cite{Minnhagen}.
Using the fact that, according to Eq. (\ref{2.2}), $-\Delta/(2\pi)$ 
is the inverse operator of the Coulomb potential $v$ and renormalizing
the particle fugacity $z$ by the (divergent) self-energy term
$\exp[\beta v(0)/2]$, the grand partition function (\ref{2.4}) 
can be turned via the Hubbard-Stratonovich transformation into
\begin{subequations} \label{2.5}
\begin{equation} \label{2.5a}
\Xi(z) = \frac{\int {\cal D}\phi \exp\left[ - S(z) \right]}{\int 
{\cal D}\phi \exp\left[ - S(0) \right]} ,
\end{equation}
where
\begin{equation} \label{2.5b}
S(z) = \int_{\mathbb{R}^2} {\rm d}^2 r
\left[ \frac{1}{16\pi} \left( \nabla\phi \right)^2 
- 2 z \cos( b \phi ) \right] , \quad b^2 = \frac{\beta}{4} 
\end{equation}
\end{subequations}
is the Euclidean action of the 2D sine-Gordon model.
Here, $\phi({\bf r})$ is a real scalar field and $\int {\cal D}\phi$
denotes the functional integration over this field.
The one- and two-particle densities are expressible as averages over
the sine-Gordon action (\ref{2.5b}) in the following way
\begin{subequations}
\begin{eqnarray}
n_q & = & z_q \langle {\rm e}^{{\rm i}qb\phi} \rangle , \label{2.6a} \\
n_{qq'}(\vert {\bf r}-{\bf r}'\vert) & = & z_q z_{q'} 
\langle {\rm e}^{{\rm i}qb\phi({\bf r})} 
{\rm e}^{{\rm i}q'b\phi({\bf r}')} \rangle . \label{2.6b} 
\end{eqnarray}
\end{subequations}
The renormalized fugacity parameter $z$ gets a precise meaning 
when one fixes the normalization of the coupled cos-field.
In the Coulomb-gas format, the short-distance behavior of the two-body
density for the oppositely charged particles is dominated by the
corresponding Boltzmann factor of the Coulomb potential:
$n_{+-}({\bf r},{\bf r}') \sim z_+ z_- 
\vert {\bf r}-{\bf r}' \vert^{-\beta}$ as   
$\vert {\bf r}-{\bf r}' \vert \to 0$.
With respect to the representation (\ref{2.6b}), in the sine-Gordon
picture this corresponds to
\begin{equation} \label{2.7}
\langle {\rm e}^{{\rm i}b\phi({\bf r})} {\rm e}^{-{\rm i}b\phi({\bf r}')}
\rangle \sim \vert {\bf r}-{\bf r}' \vert^{-4 b^2} \quad   
\mbox{as $\vert {\bf r}-{\bf r}' \vert \to 0$.}
\end{equation}
Under this short-distance (conformal) normalization, the divergent
self-energy factor disappears from statistical relations calculated
in the sine-Gordon format. 

The 2D Euclidean sine-Gordon action (\ref{2.5b}) takes its minimum
at a $\phi({\bf r})$ constant in space.
Due to a discrete symmetry $\phi\to\phi + 2\pi n/b$ 
($n$ being an integer), the action has infinitely many ground
states $\vert 0_n\rangle$ characterized by the associate expectation
values of the field $\langle\phi\rangle_n = 2\pi n/b$.
In the considered infinite-volume limit and for $b^2<1$,
these ground states become all degenerate \cite{Zamolodchikov79}.
Equivalently, the discrete $\phi$-symmetry is spontaneously broken
an so it is sufficient to develop the sine-Gordon action (\ref{2.5b})
around any one of its ground states, say the one $\vert 0_0 \rangle$
with $\langle\phi\rangle_0 = 0$.

In order to pass from the present Lagrangian formulation (\ref{2.5}) 
to the Hamiltonian one, one chooses say the $x$-direction to be
the ``Euclidean time'' and associates a Hilbert space ${\cal H}$
to any equal-time section $\{ x={\rm const}, y\in (-\infty,\infty) \}$.
 (Choosing the y-direction to be the ``Euclidean time'' 
gives equivalent quantization).
General states are vectors in ${\cal H}$ whose evolution
is governed by the Hamiltonian operator
\begin{equation} \label{2.8}
H = \int_{-\infty}^{\infty} {\rm d}y
\left[4\pi \Pi^2+ \frac{1}{16 \pi} (\partial_y \phi)^2 
+ 2 z \cos(b \phi) \right] .
\end{equation}
In the considered region $b^2<1$ with the spontaneously broken
discrete $\phi$-symmetry, the sine-Gordon field theory is massive
in the sense that ${\cal H}$ is the Fock space of multiparticle
states.
After rotation $x={\rm i}t$ to the (1+1) Minkowski time-space $(t,y)$,
these multiparticle states are interpreted as the asymptotic
``in-'' and ``out-'' scattering states (see below).

The particle spectrum of the (1+1)-dimensional sine-Gordon field
theory is the following \cite{Zamolodchikov79}.
The basic particles are the soliton $S$ and the antisoliton ${\bar S}$
which form a particle-antiparticle pair of equal masses $M$.
They correspond to specific $\phi$-configurations that interpolate
between two neighbouring ground states, say $\vert 0_0 \rangle$ and
$\vert 0_1 \rangle$.
Defining the ``topological charge'' $q$ as
\begin{equation} \label{2.9}
q = \frac{b}{2\pi} \int_{-\infty}^{\infty} {\rm d}y
\frac{\partial}{\partial y} \phi(x,y) = \frac{b}{2\pi}
\left[ \phi(x,\infty) - \phi(x,-\infty) \right] ,
\end{equation}
$q=+1$ $(-1)$ for the soliton (antisoliton). 
Due to topological reasons, the soliton and the 
antisoliton can coexist in the particle spectrum only 
in neutral pairs.
The $S-{\bar S}$ pair can create neutral $(q=0)$ bound states
$\{ B_i; i=1,2,\ldots < p^{-1} \}$; these particles are 
called ``breathers''.
Their number depends on the inverse of the parameter
\begin{equation} \label{2.10}
p = \frac{b^2}{1-b^2} \quad \left( = \frac{\beta}{4-\beta} \right) .
\end{equation}
The mass of the $B_i$-breather is given by
\begin{equation} \label{2.11}
m_i = 2 M \sin \left( \frac{\pi p}{2} i \right) ,
\end{equation}
and this breather disappears from the particle spectrum just when 
$m_i = 2 M$ (i.e. $p = 1/i$). 
Note that the breathers exist only in the stability region of the
point-like Coulomb gas $0<\beta<2$ $(0<p<1)$; the lightest
$B_1$-breather disappears just at the border $b^2=1/2$ $(\beta=2)$, 
which is the field-theoretical manifestation of the collapse phenomenon.
The $S-{\bar S}$ pair remains in the spectrum up to the
Kosterlitz-Thouless transition point $b^2=1$ ($\beta=4$) beyond
which the sine-Gordon model ceases to be massive.
In what follows, we shall restrict ourselves to the stability
region of point-like charges $\beta<2$.

Let $a\in\{ S,{\bar S}, B_i (i=1,2,\ldots < p^{-1})\}$
denote the type of the given particle and $m_a$ the corresponding
particle mass.
Since the sine-Gordon model is a relativistic field theory,
the energy $E$ and the momentum $p$ of the particle can be
parameterized as follows
\begin{equation} \label {2.12}
E_a = m_a \cosh \theta , \quad p_a = m_a \sinh \theta ,
\end{equation}
where $\theta\in (-\infty,\infty)$ is the particle rapidity.
The asymptotic $n$-particle states $\{ \vert n \rangle \}$
are generated by the ``particle creation operators'' $A_a^+(\theta)$,
\begin{equation} \label{2.13}
A_{a_1}^+(\theta_1) A_{a_2}^+(\theta_2) \cdots A_{a_n}^+(\theta_n)
\vert 0 \rangle ,
\end{equation}
where $\vert 0 \rangle \in {\cal H}$ is the ground state of the
Hamiltonian $H$ given by (\ref{2.8}).
The state (\ref{2.13}) is interpreted as an in-state if the
rapidities are ordered as $\theta_1 > \theta_2 > \cdots > \theta_n$
and as an out-state in the case of the reverse order
$\theta_1 < \theta_2 < \cdots < \theta_n$. 
The in-state basis and the out-state basis are related via 
the scattering $n\to n$ $S$-matrix.
The $2\to 2$ process is described simply by
\begin{equation} \label{2.14}
A_{a_1}^+(\theta_1) A_{a_2}^+(\theta_2) = \sum_{b_1,b_2}
S_{a_1 a_2}^{b_1 b_2}(\theta_1-\theta_2) 
A_{b_2}^+(\theta_2) A_{b_1}^+(\theta_1) . 
\end{equation}
Here, the momentum conservation demands $m_{a_1} = m_{b_1}$
and $m_{a_2} = m_{b_2}$, so that the inequalities $a_1\ne b_1$ or 
$a_2\ne b_2$ are allowed only for the soliton-antisoliton pair 
with the degenerate masses equal to $M$.
Like for any integrable field theory, the $n\to n$ $S$-matrix
of the sine-Gordon model factorizes into a product of
$n(n-1)/2$ two-particle $S$-matrices.
The two-particle $S$-matrix possesses many symmetry constraints
and its explicit form was obtained by exploring four general
axioms: the Yang-Baxter equation; a unitarity condition;
analyticity and crossing symmetry; the bootstrap principle 
\cite{Zamolodchikov79}.

The (dimensionless) specific grand potential $\omega$ of
the sine-Gordon model (\ref{2.5}), defined in the infinite-volume
limit as
\begin{equation} \label{2.15}
- \omega = \frac{1}{\vert \mathbb{R}^2\vert} \ln \Xi ,
\end{equation}
was found by using the TBA in Ref. \onlinecite{Destri}:
\begin{equation} \label{2.16}
- \omega = \frac{m_1^2}{8 \sin(\pi p)} .
\end{equation}
Here, $m_1$ is the mass of the lightest $B_1$-breather [see formula
(\ref{2.11}) taken with $i=1$] and the parameter $p$ is defined
by (\ref{2.10}).
Under the conformal normalization (\ref{2.7}), the relationship
between the fugacity $z$ and the soliton mass $M$ was established
in Ref. \onlinecite{Zamolodchikov95}:
\begin{equation} \label{2.17}
z = \frac{\Gamma(b^2)}{\pi \Gamma(1-b^2)} 
\left[ M \frac{\sqrt{\pi} \Gamma\left( (1+p)/2 \right)}{2
\Gamma(p/2)} \right]^{2-2b^2} ,
\end{equation}   
where $\Gamma$ stands for the Gamma function.
Since the total particle number density $n$ of the 2D Coulomb gas is given
by the obvious equality
\begin{equation} \label{2.18}
n = z \frac{\partial (-\omega)}{\partial z} ,
\end{equation}
Eqs. (\ref{2.16}) and (\ref{2.17}) imply the explicit density-fugacity
relationship, and consequently the complete bulk thermodynamics, 
of the 2D Coulomb gas in the whole stability region $\beta<2$
\cite{Samaj00}.
The mass $m_1$ plays the role of the inverse correlation length
of the plasma particles \cite{Samaj02a,Samaj02b}.
Using Eqs. (\ref{2.16}) - (\ref{2.18}), it is expressible as
\begin{eqnarray} 
m_1 & = & \kappa \left[ \frac{\sin(\pi p)}{\pi p} \right]^{1/2}
\nonumber \\ & = &
\kappa \left[ 1 - \frac{\pi^2}{192} \beta^2 + O(\beta^3) \right] ,
\label{2.19}
\end{eqnarray}
where
\begin{equation} \label{2.20}
\kappa = \sqrt{2\pi\beta n}
\end{equation}
is the inverse Debye length.
In the DH limit $\beta\to 0$, $m_1$ reduces to $\kappa$ as it should be.

\renewcommand{\theequation}{3.\arabic{equation}}
\setcounter{equation}{0}

\section{The 2D electrical double layer}

\subsection{The definition}
We first introduce in detail the model of interest which describes
an electrode in contact with a classical electrolyte.
Let us consider an infinite 2D space of points ${\bf r}\in \mathbb{R}^2$
defined by Cartesian coordinates $(x,y)$.
The electrode-electrolyte interface is localized at $x=0$, along the
$y$ axis.
The half-space ${\bar \Lambda} = \{ (x,y); x<0 \}$ is occupied by 
an ideal-conductor wall of dielectric constant $\epsilon_W \to \infty$, 
impenetrable to electrolyte particles.
The electrolyte is localized in the complementary half-space 
$\Lambda = \{ (x,y); x>0 \}$.
It is modeled by the classical 2D Coulomb gas of point-like 
unit $\pm$ charges.
The interface $(x=0)$ is kept at zero potential while the bulk
interior of the electrolyte $(x\to\infty)$ is assumed to be at some
electrostatic potential $\varphi$.
The non-zero potential $\varphi$ causes a splitting of 
the charge fugacities:
\begin{equation} \label{3.1}
z_{\pm} = z \exp \left( \pm \beta \varphi \right) .
\end{equation}
Alternatively, chemical potentials $\mu_+$ and $\mu_-$ can be defined by
$z_{\pm} = \exp(\beta \mu_{\pm})/\lambda^2$ where $\lambda$ is the
de Broglie thermal wavelength.
The bulk Coulomb gas is neutral \cite{Lieb}, and therefore its bulk 
properties depend only on the chemical potential combination 
$\mu = (\mu_+ + \mu_-)/2$, i.e. on $z$.
The difference $\mu_+ - \mu_-$ (or $\varphi$) is relevant only for
the surface properties of the electrolyte region close to the interface
(the electrical double layer) \cite{Jancovici86,Forrester}; 
if $\varphi\ne 0$ the electrical double layer carries some surface 
charge density.

The presence of the ideal-conductor wall is described mathematically
by charge images: the particle with charge $q$ localized in the
electrolyte region at the point ${\bf r} = ( x>0, y )$ induces 
the image with the opposite charge $q^* = -q$ localized in 
the wall region at the point ${\bf r}^* = (-x,y)$.
The interaction energy $E$ of a set of particles 
$\{ q_i, {\bf r}_i = ( x_i>0, y_i ) \}$ then consists of two parts
(see, for example, Ref. \onlinecite{Jackson}):
\begin{equation} \label{3.2}
E(\{ q_i,{\bf r}_i \}) = \sum_{i<j} q_i q_j v(\vert {\bf r}_i-{\bf r}_j \vert)
+ \frac{1}{2} \sum_{i,j} q_i q_j^* v(\vert {\bf r}_i-{\bf r}^*_j \vert) ;
\end{equation}  
the first term corresponds to direct particle-particle interactions,
while the second term describes interactions of particles with the
images of other particles and with their self-images.
The grand partition function $\Xi_{\rm bry}$ is again given by 
\begin{subequations} \label{3.3}
\begin{equation} \label{3.3a}
\Xi_{\rm bry} = \sum_{N_+=0}^{\infty} \sum_{N_-=0}^{\infty}
\frac{z_+^{N_+}}{N_+!} \frac{z_-^{N_-}}{N_-!} Q(N_+,N_-) , 
\end{equation}
where the configuration integral
\begin{equation} \label{3.3b}
Q(N_+,N_-) = \int_{\Lambda} \prod_{i=1}^N {\rm d}^2 r_i
\exp \left[ - \beta E(\{ q_i,{\bf r}_i \}) \right]
\end{equation}
\end{subequations}
is now restricted to the half-space $x>0$. 
The stability range of inverse temperatures for the surface 
thermodynamics is determined by the Coulomb interaction of 
the charged particle with its image \cite{Samaj01a}: 
the Boltzmann factor of a particle at a distance $x$ from the wall 
with its own image, proportional to $(2 x)^{-\beta/2}$, is integrable 
at small 1D distances $x$ provided that $\beta < 2$. 
Note that the bulk and surface thermodynamic stability intervals 
of $\beta$ coincide.

Due to the translational invariance of the system along the $y$ axis,
the species densities depend only on the $x$-coordinate: 
$n_{\pm}({\bf r}) \equiv n_{\pm}(x)$, $x\ge 0$.
The electrolyte is neutral in the bulk interior of the electrolyte,
i.e. it holds $\lim_{x\to\infty} n_{\pm}(x) = n_{\pm} = n/2$.
The species densities determine the particle number density
\begin{equation} \label{3.4}
n(x) = \sum_{q=\pm 1} n_q(x) = n_+(x) + n_-(x)
\end{equation}
and the charge density
\begin{equation} \label{3.5}
\rho(x) = \sum_{q=\pm 1} q n_q(x) = n_+(x) - n_-(x) .
\end{equation}
The induced (averaged) electrostatic potential $\varphi(x)$ in the
electrolyte region is related to the charge density via the Poisson equation
\begin{equation} \label{3.6}
\varphi''(x) = - 2 \pi \rho(x) .
\end{equation}
The potential satisfies the obvious boundary conditions $\varphi(0) = 0$ 
and $\varphi(\infty) = \varphi$, and the regularity requirement
(all derivatives $\varphi'$, $\varphi''$, etc. vanish) at $x\to\infty$.
Since we will be interested in the asymptotic approach of $\varphi(x)$ 
to its bulk value $\varphi$ at large distance $x$ from the electrode
surface, it is natural to introduce the quantity
\begin{equation} \label{3.7}
\delta\varphi(x) = \varphi(x) - \varphi
\end{equation} 
which vanishes as $x\to\infty$.

There are two sum rules which can be derived without solving
explicitly the boundary problem.
First, the consideration of Eq. (\ref{3.6}) in the integral
$\int_0^{\infty} {\rm d}x\, \rho(x)$ implies
\begin{equation} \label{3.8}
\int_0^{\infty} {\rm d}x\, \rho(x) = \frac{1}{2\pi} \varphi'(0) . 
\end{equation}
We note that a nonzero surface (more precisely ``line'') charge
in the electrolyte $\int_0^{\infty} {\rm d}x \rho(x)$ can appear only
in the special case of an ideal-conductor wall where this charge is exactly 
compensated by the opposite surface charge $\sigma$ of particle
images, and the system as a whole is neutral. 
Since an isolated charged line at $x=0$, carrying a uniform charge $\sigma$
per unit length, induces the electric potential such that
\begin{equation} \label{3.9}
\varphi'(0) = - 2\pi\sigma ,
\end{equation}
the sum rule (\ref{3.8}) can be understood as the neutrality-type condition
\begin{equation} \label{3.10}
\int_0^{\infty} {\rm d}x\, \rho(x) + \sigma = 0 .
\end{equation}
Second, the consideration of Eq. (\ref{3.6}) in the integral
$\int_0^{\infty} {\rm d}x\, x \rho(x)$ and the subsequent two
integrations by parts lead to 
\begin{equation} \label{3.11}
\int_0^{\infty} {\rm d}x\, x \rho(x) = \frac{\varphi}{2\pi} ,
\end{equation}
i.e. the dipole moment of the charge density is related to
the potential difference across the electrical double layer.
This relation follows from elementary macroscopic electrostatics
\cite{Jackson}.

At small distance from the wall $x\to 0$, the species densities
are determined by the Boltzmann factors of the corresponding
particle with its image \cite{Samaj01a}:
\begin{equation} \label{3.12}
n_{\pm}(x) \sim \frac{z_{\pm}}{(2 x)^{\beta/2}} , \quad x\to 0 . 
\end{equation}
Consequently,
\begin{subequations} \label{3.13}
\begin{eqnarray}
n(x) & \displaystyle{\mathop{\sim}_{x\to 0}} & 2 z
\frac{\cosh (\beta\varphi)}{(2 x)^{\beta/2}} , \label{3.13a} \\
\rho(x) & \displaystyle{\mathop{\sim}_{x\to 0}} & 2 z
\frac{\sinh (\beta\varphi)}{(2 x)^{\beta/2}} . \label{3.13b}
\end{eqnarray}
\end{subequations}

\subsection{The boundary sine-Gordon representation}
The considered particle-image system is mappable onto a boundary
sine-Gordon theory \cite{Samaj01a}.
In particular, the grand partition function (\ref{3.3}) can be
written as
\begin{subequations} \label{3.14}
\begin{equation} \label{3.14a}
\Xi_{\rm bry}(z) = \frac{\int {\cal D}\phi 
\exp\left( - S_{\rm bry}(z) \right)}{\int 
{\cal D}\phi \exp\left( - S_{\rm bry}(0) \right)} ,
\end{equation}
where
\begin{equation} \label{3.14b}
S_{\rm bry}(z) = \int_{x>0} {\rm d}^2 r
\left[ \frac{1}{16\pi} \left( \nabla\phi \right)^2 
- 2 z \cos( b \phi ) \right] , \quad b^2 = \frac{\beta}{4} 
\end{equation}
\end{subequations}
is the 2D Euclidean action of the boundary sine-Gordon model defined
in the half-space $x\ge 0$ and the real scalar field $\phi({\bf r})$
fulfills the following Dirichlet conditions at the $x=0$ boundary:
\begin{equation} \label{3.15}
\phi(x=0,y) = \phi_0 = - 4 {\rm i}  b \varphi .
\end{equation}
The fact that the boundary value of the field, $\phi_0$, is a pure
imaginary number makes no problem: as is usual in the field theory,
first one expresses the quantities of interest as functions of real
$\phi_0$ and then analytically continues the obtained results to
complex values of $\phi_0$.
This procedure was successfully applied to the calculation of
the surface tension (i.e. the surface part of the grand potential)
for the present model \cite{Samaj01a}.
$\phi_0$ will usually appear in the combination
\begin{equation} \label{3.16}
\eta = - \frac{\phi_0}{2 b} = 2 {\rm i}  \varphi .
\end{equation} 
Within the formalism developed in Ref. \onlinecite{Samaj01a},
the one-particle densities $n_{\pm}(x)$ in the electrolyte region
$x\ge 0$ are expressible as averages over the boundary sine-Gordon
action (\ref{3.14b}) with Dirichlet boundary conditions
(\ref{3.15}) as follows
\begin{equation} \label{3.17}
n_{\pm}(x) = z \langle {\rm e}^{\pm {\rm i} b \phi(x,0)} 
\rangle_{\rm bry} .
\end{equation}
Here, regarding the $y$-invariance of the boundary mean values
$\langle {\rm e}^{\pm {\rm i} b \phi({\bf r})} \rangle_{\rm bry}$,
we have set $y=0$ for simplicity.
Thus, the particle number density (\ref{3.4}) and the charge density  
(\ref{3.5}) read
\begin{eqnarray} 
n(x) & = & z \left[ \langle {\rm e}^{{\rm i} b \phi(x,0)} 
\rangle_{\rm bry} + \langle {\rm e}^{- {\rm i} b \phi(x,0)} 
\rangle_{\rm bry} \right] , \label{3.18}
\\
\rho(x) & = & z \left[ \langle {\rm e}^{{\rm i} b \phi(x,0)} 
\rangle_{\rm bry} - \langle {\rm e}^{- {\rm i} b \phi(x,0)} 
\rangle_{\rm bry} \right] , \label{3.19}
\end{eqnarray}
respectively.

The specific case of Dirichlet boundary conditions
in the semi-infinite sine-Gordon model does not spoil 
the integrability property of the bulk theory \cite{Ghoshal1}.
In passing from the Lagrangian formulation (\ref{3.14}) to a Hamiltonian one,
contrary to the bulk case, we have two different choices considering
either $x$ or $y$ as the Euclidian time. 

If $y$ is taken to be the Euclidean time, then the boundary is in space, and
a boundary Hilbert space ${\cal H}_B$ is associated to any time slices 
$\{ x \in (0,\infty),y={\rm const} \}$. The time evolution is governed by 
the Hamiltonian  
\begin{equation} \label{3.20}
H_B = \int_{0}^{\infty} {\rm d}y
\left[4\pi \Pi^2+ \frac{1}{16 \pi} (\partial_y \phi)^2 
+ 2 z \cos(b \phi) \right]\quad ,
\end{equation}
where the boundary condition, (\ref{3.15}),  is satisfied. 
The boundary Hilbert space consists of boundary bound states, \cite{Mattsson},
and bulk multiparticle states: 
\begin{equation} \label{3.21}
A_{a_1}^+(\theta_1) A_{a_2}^+(\theta_2) \cdots A_{a_n}^+(\theta_n)
\vert 0 \rangle _B ,
\end{equation}
where $\vert 0 \rangle _B \in {\cal H}_B$ is the ground state of the
Hamiltonian $H_B$ given by (\ref{3.20}).
The state (\ref{3.21}) is interpreted as an in-state if the
rapidities are ordered as $0>\theta_1 > \theta_2 > \cdots > \theta_n$
and as an out-state in the case of the reverse order
$0 < \theta_1 < \theta_2 < \cdots < \theta_n$. 
The in-state basis and the out-state basis are related via 
the reflection $n\to n$ $R$-matrix. This matrix factorizes into the product of
pairwise scatterings (\ref{2.14}) and of, $R_a^b(\theta)$-s, the individual
one-particle  reflections
$A^+_a(-\theta) \to A^+_b(\theta)$ ($\theta$ positive) off the boundary.
Owing to the energy conservation, $R_a^b(\theta)$ vanishes if $m_a\ne m_b$.
The $R$-amplitudes were obtained explicitly for the soliton/antisoliton 
pair in Ref. \onlinecite{Ghoshal1} and for the breathers in Ref.
\onlinecite{Ghoshal2}.  
We shall need the Dirichlet $R$-amplitudes for the lowest $B_1$- and 
$B_2$-breathers; with the notation 
$R_{B_j}^{B_j}(\theta) \equiv R_B^{(j)}(\theta)$, one has explicitly
\begin{subequations} \label{3.22}
\begin{equation} \label{3.22a}
R_B^{(1)}(\theta) = \frac{
\left( \frac{1}{2} \right) \left( \frac{p}{2}+1 \right) 
\left( \frac{\eta p}{\pi} - \frac{1}{2} \right)}{
\left( \frac{p}{2}+\frac{3}{2} \right) 
\left( \frac{\eta p}{\pi} + \frac{1}{2} \right)}
\end{equation}
and
\begin{eqnarray} 
R_B^{(2)}(\theta) & = & \frac{
\left( \frac{1}{2} \right) \left( \frac{p}{2}+1 \right)
\left( p+1 \right) \left( \frac{p}{2} \right) }{
\left( \frac{p}{2}+\frac{3}{2} \right)^2 
\left(  p + \frac{3}{2} \right) } \nonumber \\ & & \times
\frac{ \left( \frac{\eta p}{\pi} - \frac{1}{2} -\frac{p}{2}\right)
\left( \frac{\eta p}{\pi} - \frac{1}{2} +\frac{p}{2}\right)}{ 
\left( \frac{\eta p}{\pi} + \frac{1}{2} -\frac{p}{2}\right)
\left( \frac{\eta p}{\pi} + \frac{1}{2} +\frac{p}{2}\right) } , \label{3.22b}
\\ \nonumber
\end{eqnarray}
\end{subequations}
where we used the symbol
\begin{equation} \label{3.23}
(x) = \frac{\sinh \left( 
\frac{\theta}{2} + \frac{{\rm i}\pi x}{2} \right)}
{\sinh \left( 
\frac{\theta}{2} - \frac{{\rm i}\pi x}{2} \right)} .
\end{equation}
The reflection amplitudes of breathers have simple poles at 
the imaginary rapidity $\theta = {\rm i}\pi/2$.
In the particular case of the first two breathers (\ref{3.22}),
one finds that
\begin{equation} \label{3.24}
R_B^{(j)}(\theta) \sim \frac{{\rm i} g_j^2}{2 \theta - {\rm i}\pi} ,
\quad j=1,2 
\end{equation} 
where the ``boundary couplings'' $g_1$ and $g_2$ are extracted in the form
\begin{subequations}
\begin{eqnarray} 
g_1 & = & 2 \tan \left( \frac{p\eta}{2} \right)
\left[ \frac{1+ \cos\left( \frac{p\pi}{2}\right)
- \sin\left( \frac{p\pi}{2}\right)}{1-\cos\left( \frac{p\pi}{2}\right)
+ \sin\left( \frac{p\pi}{2}\right)} \right]^{1/2} 
,\phantom{rrrrr} \label{3.25a} \\
g_2 & = & \frac{2 \tan\left( \frac{p\pi}{4} - \frac{p\eta}{2}
 \right) \tan \left( \frac{p\pi}{4} + \frac{p\eta}{2} \right)}{
\tan \left( \frac{p\pi}{4} \right) \left[
\tan \left( \frac{p\pi}{2} \right)
\tan \left( \frac{\pi}{4} + \frac{p\pi}{2} \right) \right]^{1/2}} .
\label{3.25b} \\ \nonumber
\end{eqnarray}
\end{subequations} 

In the alternative Hamiltonian description one can 
take $x$ to be the Euclidean time and associate with 
any equal-time section $\{ x={\rm const}, y\in (-\infty,\infty) \}$ 
the same Hilbert space ${\cal H}$ as in the bulk theory.
The Hamiltonian operator is now given by Eq. (\ref{2.8}).
The boundary at $x=0$ appears as the initial condition described
by the boundary state $\vert B\rangle \in {\cal H}$.
In the 1+1 Minkowski space-time, $\vert B\rangle$ can be written
as a superposition of the bulk asymptotic states (\ref{2.13}):
\begin{eqnarray} 
\hspace{-2cm} \vert B \rangle & = & \exp \Bigg\{ \sum_a {\tilde g}_a A_a^+(0)+
\nonumber \\ & &
\int_0^{\infty} \frac{{\rm d}\theta}{2\pi} 
\sum_{(a,b)} K^{ab}(\theta) A^+_a(-\theta) A^+_b(\theta) 
\Bigg\} \vert 0 \rangle . \label{3.26} \\ \nonumber
\end{eqnarray} 
The amplitude $K^{ab}(\theta)$ is related to the reflection matrix as
\begin{equation} \label{3.27}
K^{ab}(\theta) = R_{{\bar a}}^b\left( \frac{{\rm i}\pi}{2} -\theta
\right) ,
\end{equation}
where ${\bar a}$ denotes the antiparticle of the particle $a$
$({\bar B}_j = B_j)$.
Ghoshal and Zamolodchikov \cite{Ghoshal1} identified ${\tilde g}_a$
with the boundary coupling $g_a$, but Dorey et al \cite{Dorey}
found in the case of the Lee-Yang model the relation
\begin{equation} \label{3.28}
{\tilde g}_a = g_a/2 .
\end{equation}
The strong evidence that this formula extends to any boundary 
(1+1)-dimensional quantum field theory, and in particular to 
the sine-Gordon one, was presented later in Ref. \onlinecite{Bajnok05}.

\renewcommand{\theequation}{4.\arabic{equation}}
\setcounter{equation}{0}

\section{Mean-field theories}

\subsection{The Debye-H\"uckel limit}
The position-dependent species densities in the electrical
double layer can be evaluated systematically via an expansion
in powers of $\beta$ around the DH high-temperature
limit $\beta\to 0$.
The technique is based on a Mayer expansion of the free energy 
with series-renormalized bonds between each couple of field circles; 
for a detailed description of the method see Ref. \onlinecite{Samaj01a}.

In the lowest expansion order, the species are considered to be
constant in the electrolyte region, $n_{\pm}(x) = n/2$ $(x\ge 0)$.
Under this assumption, the renormalized bond is given by
$K = K^{(0)}$ with
\begin{equation} \label{4.1}
K^{(0)}({\bf r}_1,{\bf r}_2) = - \beta K_0(\kappa r_{12})
+ \beta K_0(\kappa r^*_{12}) .
\end{equation}
Here, $\kappa$ is the inverse Debye length (\ref{2.20}),
$K_0$ is the modified Bessel function of second kind,
$r_{12} = \vert {\bf r}_1-{\bf r}_2 \vert$ and
$r_{12}^* = \vert {\bf r}_1-{\bf r}^*_2 \vert
= \vert {\bf r}^*_1 - {\bf r}_2 \vert$.
The first correction to the constant species densities can be
obtained iteratively by inserting the lowest-order $K^{(0)}$
into a basic generating formula for the density profiles, 
with the result
\begin{equation} \label{4.2}  
n_{\pm}(x) = \frac{n}{2} \exp \left\{ \pm \beta 
[\varphi - \varphi(x)] + \frac{\beta}{2} K_0(2\kappa x) \right\} .
\end{equation}
Here, $\varphi(x)$ is the electric potential induced by the charge
density $\rho(x)$ via the Poisson equation (\ref{3.6}).

Expanding the exponential in (\ref{4.2}) to order $\beta$ 
gives for the charge density
\begin{equation} \label{4.3}
\rho(x) = \beta n \left[ \varphi - \varphi(x) \right] .
\end{equation} 
Inserting this into Eq. (\ref{3.6}) and considering the boundary 
conditions for the electric potential, one gets
\begin{subequations} \label{4.4}
\begin{eqnarray}
\delta\varphi_{\rm DH}(x) & = & 
- \varphi \exp ( -\kappa x ) , \label{4.4a} \\
\rho_{\rm DH}(x) & = & \beta n \varphi \exp ( -\kappa x ) , \label{4.4b} 
\end{eqnarray}
\end{subequations} 
where the deviation of the electric potential from its bulk value
is defined by Eq. (\ref{3.7}).
The surface DH ``image-charge'' $\sigma_{\rm DH}$, defined by the couple of
equivalent Eqs. (\ref{3.9}) and (\ref{3.10}), is obtained in the form
\begin{equation} \label{4.5}
\sigma_{\rm DH} = - \frac{\kappa \varphi}{2\pi} .
\end{equation}
In terms of $\sigma_{\rm DH}$, the relation (\ref{4.4a}) is written as
\begin{equation} \label{4.6}
\delta \varphi_{\rm DH}(x) = \frac{2\pi \sigma_{\rm DH}}{\kappa} 
\exp(-\kappa x) .
\end{equation}

Expanding the exponential in (\ref{4.2}) to order $\beta$, 
the particle number density reads
\begin{equation} \label{4.7}
n_{\rm DH}(x) = n + \delta n_{\rm DH}(x), \quad 
\delta n_{\rm DH}(x) = \frac{\beta n}{2} K_0(2\kappa x) .
\end{equation}
At large distances $x$ from the wall,
\begin{equation} \label{4.8}
\delta n_{\rm DH}(x) \sim 
\frac{\beta n}{2} \left( \frac{\pi}{4\kappa x} 
\right)^{1/2} \exp ( - 2\kappa x ) , \quad x\to\infty .
\end{equation}

\subsection{The leading high-temperature correction}
The next iteration for $n_{\pm}(x)$ is obtained by using the DH density
(\ref{4.7}) in the basic equation for the renormalized bond $K$ 
and treating $\delta n(x)$ as a perturbation.
Now $K = K^{(0)} + K^{(1)}$, where $K^{(0)} \propto \beta$ is defined
by Eq. (\ref{4.1}) and $K^{(1)} \propto \beta^2$ is given,
to first order in the density perturbation, by the integral equation
\begin{equation} \label{4.9}
K^{(1)}({\bf r}_1,{\bf r}_2) = \int_{x_3>0} {\rm d}^2 r_3\, 
K^{(0)}({\bf r}_1,{\bf r}_3) \delta n(x_3) K^{(0)}({\bf r}_3,{\bf r}_2) .
\end{equation}

Taking also $K^{(1)}$ into account, instead of (\ref{4.2}) we have
\begin{equation} \label{4.10}
n_{\pm}(x) = \frac{n}{2} \exp \left\{ \mp \beta \delta\varphi(x) 
+ \frac{\beta}{2} K_0(2\kappa x) + \frac{1}{2} K^{(1)}(x) \right\} ,
\end{equation}
where $K^{(1)}({\bf r},{\bf r})$, with ${\bf r}$ having the coordinate $x$,
is renamed $K^{(1)}(x)$.
Expanding the exponential in (\ref{4.10}) up to order $\beta^2$,
one obtains 
\begin{eqnarray}
\rho(x) & = & - \beta n \delta\varphi(x) - \frac{\beta^2 n}{2}
K_0(2\kappa x) \delta\varphi(x) , \label{4.11} \\
\delta n(x) & = & \frac{\beta n}{2} K_0(2\kappa x) + 
\frac{n}{2} K^{(1)}(x) \nonumber \\
& & + \frac{\beta^2 n}{2} \left[ \delta\varphi(x) \right]^2 
+ \frac{\beta^2 n}{8} [K_0(2\kappa x)]^2 . \label{4.12}
\\ \nonumber
\end{eqnarray}

Inserting the charge density (\ref{4.11}) into the Poisson equation 
(\ref{3.6}) and considering the $\beta$-expansion in the form
\begin{equation} \label{4.13}
\delta\varphi(x) = - \varphi {\rm e}^{-\kappa x}
+ \frac{\beta\varphi}{2} f(\kappa x) ,
\end{equation} 
the unknown $f$-function is determined by the ordinary differential
equation
\begin{equation} \label{4.14}
f''(x) - f(x) = - {\rm e}^{-x} K_0(2x)
\end{equation}
with the zero boundary conditions $f(0) = f(\infty) = 0$.
In terms of the 1D Green function
\begin{equation} \label{4.15}
G(x,x') = - {\rm e}^{-x_>} \sinh (x_<) 
\end{equation}
with $x_< = \min \{ x,x' \}$ and $x_> = \max \{ x,x' \}$,
the solution of Eq. (\ref{4.14}) reads
\begin{equation} \label{4.16}
f(x) = \int_0^{\infty} {\rm d}x'\, G(x,x')
\left[ - {\rm e}^{-x'} K_0(2 x') \right] .
\end{equation}
At large $x$, $f(x)$ behaves like ${\rm e}^{-x} [(\pi/2)-1]/4$.
Consequently,
\begin{subequations} \label{4.17}
\begin{eqnarray}
\delta\varphi(x) & \displaystyle{\mathop{\sim}_{x\to\infty}} &  
- \varphi \left[ 1 - \frac{\beta}{8} \left( \frac{\pi}{2} -1 \right) 
\right] {\rm e}^{-\kappa x} , \label{4.17a} \\ 
\rho(x) & \displaystyle{\mathop{\sim}_{x\to\infty}} &  
\beta n \varphi \left[ 1 - \frac{\beta}{8} \left( \frac{\pi}{2} -1 \right) 
\right] {\rm e}^{-\kappa x} . \label{4.17b}
\end{eqnarray}
\end{subequations}
These formulae provide the pure exponential asymptotic decay of 
the DH results (\ref{4.4}), in agreement with the concept of 
renormalized charge \cite{Belloni,Diehl,Trizac2,Aubouy}.
The renormalized image charge $\sigma_{\rm ren}$, defined in analogy
with Eq. (\ref{4.6}) as
\begin{equation} \label{4.18}
\delta \varphi(x) \mathop{\sim}_{x\to\infty} 
\frac{2\pi \sigma_{\rm ren}}{\kappa} \exp(-\kappa x) ,
\end{equation}
has the following weak-coupling expansion
\begin{equation} \label{4.19}
\sigma_{\rm ren} = - \frac{\kappa \varphi}{2\pi} \left[ 1 - 
\frac{\beta}{8}\left( \frac{\pi}{2} - 1 \right) + O(\beta^2) \right] .
\end{equation}
From Eq. (\ref{4.16}) one gets that $f'(0) = 1/2$.
Combining then Eqs. (\ref{3.9}) and (\ref{4.13}), 
the leading $\beta$-correction to the surface ``image-charge''
(\ref{4.5}) reads
\begin{equation} \label{4.20}
\sigma = - \frac{\kappa \varphi}{2\pi} \left[ 1 + \frac{\beta}{4} 
+ O(\beta^2) \right] .
\end{equation} 

The problem of the number density deviation from its bulk value,
$\delta n(x)$ given by Eq. (\ref{4.12}), is more complicated
because the function $K^{(1)}(x)$ is only defined implicitly as 
the solution of the integral equation (\ref{4.9}).
It can be shown after lengthy algebra that
\begin{equation} \label{4.21}
K^{(1)}(x) \sim \frac{\pi^2 \beta^2}{16} \exp ( - 2\kappa x ) 
\quad \mbox{as $x\to\infty$.}  
\end{equation}
Thus, at large distances from the wall,
\begin{equation} \label{4.22}
\delta n(x) \mathop{\sim}_{x\to\infty} \frac{\beta^2 n}{2}
\left( \varphi^2 + \frac{\pi^2}{16} \right) \exp ( -2 \kappa x ) .
\end{equation}
Note that this asymptotic behavior differs fundamentally from,
and is superior to, that obtained in the DH limit (\ref{4.8}). 
We conclude that, in contrast to the charge density, 
the DH theory does not provide an adequate 
description of the large-distance decay of the particle 
number density to its bulk value. 

\subsection{The nonlinear Poisson-Boltzmann theory}
In the nonlinear PB theory, one keeps the deviation of the
electrostatic potential from its bulk value, $\varphi(x)-\varphi$,
in the exponential form (\ref{4.2}).
The DH expression for the charge density (\ref{4.3}) then takes
the nonlinear form
\begin{equation} \label{4.23}
\rho(x) = n \sinh \left\{ \beta \left[ \varphi - \varphi(x) 
\right] \right\} .
\end{equation}
The corresponding Poisson equation
\begin{equation} \label{4.24}
\varphi''(x) = - 2 \pi n \sinh \left\{ 
\beta \left[ \varphi - \varphi(x) \right] \right\}
\end{equation}
is subject to the obvious boundary conditions $\varphi(0)=0$ and
$\varphi(\infty) = \varphi$.
We recall the well-known fact that, with the identifications
$\phi_{\rm st}(x) = - 4 {\rm i} b \varphi(x)$ and $z=n/2$,
the nonlinear PB equation (\ref{4.24}) corresponds to the static
equation ($\phi$ does not depend on ``time'' $y$) of the
``classical'' variational treatment of the boundary sine-Gordon action 
(\ref{3.14b}), $\delta S_{\rm bry}[\phi_{\rm st}(x)] = 0$.

For the present semi-infinite geometry, the solution of the nonlinear
PB equation (\ref{4.24}) can be derived explicitly:
\begin{equation} \label{4.25}
\delta\varphi(x) = - \frac{2}{\beta} \ln \left[
\frac{1+{\rm e}^{-\kappa x} \tanh (\beta\varphi/4)}{1-
{\rm e}^{-\kappa x} \tanh (\beta\varphi/4)} \right] .
\end{equation}
At asymptotically large distance $x$ from the wall,
\begin{equation} \label{4.26}
\delta\varphi(x) \mathop{\sim}_{x\to\infty} - \frac{4}{\beta}
\tanh \left( \frac{\beta\varphi}{4} \right) 
{\rm e}^{-\kappa x} .
\end{equation}
This behavior is again in full agreement with the idea of renormalized
charge.
The renormalized image charge $\sigma_{\rm ren}$, defined by
Eq. (\ref{4.18}), reads
\begin{equation} \label{4.27}
\sigma_{\rm ren} = - \frac{2\kappa}{\pi\beta} 
\tanh \left( \frac{\beta\varphi}{4} \right) .
\end{equation}
In the $\beta\to 0$ limit and for a finite $\varphi$, 
the DH result (\ref{4.5}) is reproduced as it should be.
On the other hand, the leading $\beta$-correction to the DH result
in Eq. (\ref{4.19}) is not reproduced.
We shall show in the next section that the nonlinear formula (\ref{4.27})
describes correctly the scaling regime of limits $\beta\to 0$ and 
$\varphi\to\infty$ with the product $\beta\varphi$ being finite. 

The surface ``image-charge'' $\sigma$, defined by either (\ref{3.9})
or (\ref{3.10}), is obtained as follows
\begin{equation} \label{4.28}
\sigma = - \frac{\kappa}{\pi\beta} 
\sinh \left( \frac{\beta\varphi}{2} \right)
\end{equation}
For a nonzero $\beta$ and in the limit of the infinite $\sigma$-charge,
which is equivalent in view of Eq. (\ref{4.28}) to the limit
$\varphi\to\infty$, the renormalized $\sigma_{\rm ren}$ (\ref{4.27}) 
saturates at the finite value
\begin{equation} \label{4.29}
\sigma_{\rm ren}^* = - \frac{2\kappa}{\pi\beta} .
\end{equation}
More generally, for a nonzero $\beta$ and in the limit
$\varphi\to\infty$, Eq. (\ref{4.25}) reduces to
\begin{equation} \label{4.30}
\delta\varphi^*(x) = - \frac{2}{\beta}
\ln \left[ \frac{1+\exp(-\kappa x)}{1-\exp(-\kappa x)} \right] ,
\end{equation}
i.e. the electric potential deviation from its bulk value saturates at 
a finite value in every point of the electrolyte region $x>0$,
in accordance with the hypothesis of the potential saturation \cite{Tellez}.
Note that the potential saturation is a pure nonlinear effect: there is
no saturation in the DH relation (\ref{4.4a}).

\renewcommand{\theequation}{5.\arabic{equation}}
\setcounter{equation}{0}

\section{Asymptotic charge and number density profiles for $\beta<2$}

\subsection{Boundary one-point functions: general formalism}
Let us consider a general integrable 2D boundary field theory, 
defined in the half-space $x>0$ and possessing an integrable 
boundary condition at $x=0$.
For the time being, the spectrum of the corresponding bulk theory 
is supposed to contain only one particle of mass $m$.
We aim at calculating formally the mean values
$\langle {\cal O}(x,0) \rangle_{\rm bry}$ (due to the translational
$y$-invariance, $y$ is set to 0 for simplicity) of some, as yet
unspecified, local field operator ${\cal O}$.
As is explained in section III.B, in the $1+1$ Minkowski
$x$-time and $y$-space, the boundary condition at $x=0$ acts
as the initial-time condition described by the boundary state
\begin{equation} \label{5.1}
\vert B \rangle = \exp \left\{ \frac{g}{2} A^+(0) +
\int_0^{\infty} \frac{{\rm d}\theta}{2\pi} K(\theta) 
A^+(-\theta) A^+(\theta) \right\} \vert 0 \rangle
\end{equation}
which belongs to the bulk Hilbert space ${\cal H}$.
Thus,
\begin{equation} \label{5.2}
\langle {\cal O}(x,0) \rangle_{\rm bry} = 
\frac{\langle 0 \vert {\cal O}(x,0)
\vert B \rangle}{\langle 0\vert B \rangle} ;  
\end{equation}
note that the normalization is $\langle 0\vert B \rangle = 1$.

A systematic expansion for the one-point function (\ref{5.2})
can be obtained by using a complete system of the bulk $n$-particles
states $\{ \vert n\rangle \}$ forming ${\cal H}$ as follows
\begin{equation} \label{5.3}
\langle 0 \vert {\cal O}(x,0) \vert B \rangle =
\sum_{n=0}^{\infty} \langle 0 \vert {\cal O}(x,0) \vert n \rangle
\langle n \vert B \rangle .
\end{equation}
In the rapidity representation, this formula reads
\begin{eqnarray} \label{5.4} 
\langle 0 \vert {\cal O}(x,0) \vert B \rangle = & &   \\
& & \hspace{-2.2cm} \langle 0 \vert {\cal O}(x,0) \vert 0 \rangle 
+ \int_{-\infty}^{\infty} \frac{{\rm d}\theta}{2\pi}
\langle 0 \vert {\cal O}(x,0) \vert \theta \rangle
\langle \theta \vert B \rangle \nonumber \\
& & \hspace{-2.3cm} +\int_{-\infty}^{\infty} \frac{{\rm d}\theta_1}{2\pi}
\int_{\theta_1}^{\infty} \frac{{\rm d}\theta_2}{2\pi}
\langle 0 \vert {\cal O}(x,0) \vert \theta_1,\theta_2 \rangle
\langle \theta_1,\theta_2 \vert B \rangle + ... , \nonumber 
\end{eqnarray}
where $\vert\theta\rangle = A^+(\theta) \vert 0 \rangle$,
$\vert\theta_1,\theta_2\rangle = A^+(\theta_1) A^+(\theta_2) 
\vert 0 \rangle$, etc.

The matrix elements $\langle\theta\vert B\rangle$, 
$\langle\theta_1,\theta_2\vert B\rangle$, etc., which depend
on the given boundary condition, can be calculated by using 
the explicit form of the boundary state (\ref{5.1}).
The normalization condition for the two-particle scattering
\begin{equation} \label{5.5}
A(\theta_1) A^+(\theta_2) = 2\pi \delta(\theta_1-\theta_2)
+ S(\theta_2-\theta_1) A^+(\theta_2) A(\theta_1) 
\end{equation}
implies that 
\begin{equation} \label{5.6}
\langle\theta \vert B\rangle \equiv \langle 0 \vert A(\theta) B\rangle 
= \frac{g}{2} 2\pi \delta(\theta) .
\end{equation}
Analogously, we have
\begin{eqnarray} 
\langle\theta_1,\theta_2 \vert B\rangle & \equiv &
\langle 0 \vert A(\theta_2) A(\theta_1) B\rangle  \\
& = & \int_0^{\infty} \frac{{\rm d}\theta}{2\pi} K(\theta)
(2\pi)^2 \delta(\theta_1+\theta) \delta(\theta_2-\theta) ,
\nonumber 
\phantom{nnn} \label{5.7} 
\end{eqnarray}
and so on.

The matrix elements $\langle 0\vert{\cal O}(x,0)\vert\theta\rangle$,
$\langle 0\vert{\cal O}(x,0)\vert\theta_1,\theta_2\rangle$, etc., which
do not depend on the given boundary condition, are known as the
bulk multi-particle form factors.
Their $x$-dependence can be factorized out by means of a translation
on the operator ${\cal O}(x,0)$ \cite{Smirnov}:
\begin{subequations} \label{5.8}
\begin{eqnarray}
\langle 0\vert{\cal O}(x,0)\vert\theta\rangle & = &
{\rm e}^{-m x \cosh \theta} F_1 , \label{5.8a} \\
\langle 0\vert{\cal O}(x,0)\vert\theta_1,\theta_2\rangle & = &
{\rm e}^{-m x (\cosh \theta_1 + \cosh \theta_2)} 
F_2(\theta_1-\theta_2) , \nonumber \\ \label{5.8b} 
\end{eqnarray}
\end{subequations}
and so on.
The bulk form factors $F_1 = \langle 0\vert{\cal O}(0,0)\vert\theta\rangle$,
$F_2(\theta_1-\theta_2) = 
\langle 0\vert{\cal O}(0,0)\vert\theta_1,\theta_2\rangle$, etc.
can be obtained explicitly in an axiomatic way, similarly to the  
case of the two-particle scattering $S$-matrix.

Finally, using the formulas (\ref{5.6})-(\ref{5.8}) in the expansion
(\ref{5.4}), the form-factor representation of the one-point function
reads
\begin{eqnarray} \label{5.9}
\langle {\cal O}(x,0) \rangle_{\rm bry} 
= \langle 0 \vert {\cal O}(x,0) \vert 0 \rangle 
+ \frac{g}{2} F_1 {\rm e}^{-m x} & &  \\ 
& & \hspace{-5.5cm}
+ \int_0^{\infty} \frac{{\rm d}\theta}{2\pi} K(\theta)
F_2(-2\theta) {\rm e}^{-2 m x \cosh \theta} +
O\left( {\rm e}^{-3mx} \right) . \nonumber 
\end{eqnarray}
This formula is particularly useful for large distances $x$ from the
wall since it provides a systematic large-distance expansion.
The first term is nothing but the bulk expectation value of 
the operator ${\cal O}$, $\langle {\cal O} \rangle$, which is 
indeed dominant in the limit $x\to\infty$.
The leading correction term comes from the one-particle state, 
with the particle mass $m$ playing the role of 
the inverse correlation length.
The next-to-leading correction term decays at large $x$ as
$\exp(-2mx)$ multiplied by an inverse-power of $m x$, and so on.

The extension of the formula (\ref{5.9}) to general integrable 2D field
theories with many-particle spectrum $\{ a\}$ is straightforward.
We only write down the final result:
\begin{eqnarray} \label{5.10}
\langle {\cal O}(x,0) \rangle_{\rm bry} 
= \langle {\cal O} \rangle 
+ \sum_a \frac{g_a}{2} F^{(a)}_1 {\rm e}^{-m_a x} & & \\ 
& &\hspace{-6cm} + \int_0^{\infty} \frac{{\rm d}\theta}{2\pi} 
\sum_{\substack{(a,b) \\ m_a = m_b}} K^{ab}(\theta)
F^{(ab)}_2(-2\theta) {\rm e}^{-2 m_a x \cosh \theta} +
\cdots . \nonumber
\end{eqnarray}

\subsection{The electrical double layer}
Since the number (\ref{3.18}) and the charge (\ref{3.19}) density 
profiles are determined by the boundary mean values of the exponential
field, $\langle {\rm e}^{{\rm i}q b \phi(x,0)} \rangle_{\rm bry}$ 
$(q=\pm 1)$, the operator of interest is
\begin{equation} \label{5.11}
{\cal O}_q(x,0) = \exp\left\{ {\rm i} q b \phi(x,0) \right\} ,
\quad q=\pm 1 .
\end{equation}

In the spectrum of the 2D bulk sine-Gordon theory, the first two
lightest neutral particles are the $B_1$- and $B_2$-breathers with
the corresponding masses (see Eq. (\ref{2.11}))
\begin{subequations} 
\begin{eqnarray}
m_1 & = & 2 M \sin \left( \frac{\pi p}{2} \right) ,
\quad p<1 , \label{5.12a} \\
m_2 & = & 2 M \sin (\pi p) ,
\quad p < \frac{1}{2} . \label{5.12b}
\end{eqnarray}
\end{subequations}
The mass $m_2\le 2 m_1$, the equality taking place in the DH limit $p\to 0$.
Consequently, when applying the large-$x$ expansion (\ref{5.10}) to 
the boundary sine-Gordon theory with Dirichlet boundary conditions 
(\ref{3.15}), we have to consider the following leading terms
\begin{eqnarray} \label{5.13}
\langle {\cal O}_q(x,0) \rangle_{\rm bry} = \langle {\cal O} \rangle \\
& & \hspace{-2.5cm} + \frac{g_1}{2} F_1^{(1)}(q) {\rm e}^{- m_1 x}
+ \frac{g_2}{2} F_1^{(2)}(q) {\rm e}^{- m_2 x}   \nonumber \\ 
& & \hspace{-2.5cm}
+ \int_0^{\infty} \frac{{\rm d}\theta}{2\pi} K^{11}(\theta)
F_2^{(11)}(-2\theta;q) {\rm e}^{-2 m_1 x \cosh \theta}
+ \cdots . \nonumber
\end{eqnarray} 
Here, the boundary couplings $g_1$ and $g_2$ are given by Eqs. (\ref{3.25a})
and (\ref{3.25b}), respectively, and 
$K^{11}(\theta) = R_B^{(1)}(\frac{{\rm i}\pi}{2}-\theta)$
where the reflection amplitude $R_B^{(1)}$ is given by Eq. (\ref{3.22a}).

Our preliminary task is to write down the bulk form factors in 
Eq. (\ref{5.13}) for the exponential operator (\ref{5.11}).
We start with the $B_1$ form factors which have been obtained in 
Ref. \onlinecite{Lukyanov}.
The one-particle $B_1$ form factor reads
\begin{equation} \label{5.14}
F_1^{(1)}(q) = - {\rm i} q \lambda 
\langle {\rm e}^{{\rm i}q b \phi} \rangle ,
\qquad q = \pm 1 , 
\end{equation}
where the parameter $\lambda$ is defined by
\begin{equation} \label{5.15}
\lambda = 2 \cos\left( \frac{p\pi}{2} \right)
\left[ 2 \sin\left( \frac{p\pi}{2} \right) \right]^{1/2}
\exp\left( - \int_0^{p\pi} \frac{{\rm d}t}{2\pi} \frac{t}{\sin t} \right) .
\end{equation}
The two $B_1$-breathers form factor reads
\begin{equation} \label{5.16}
F_2^{(11)}(\theta;q) = - \lambda^2 R(\theta) 
\langle {\rm e}^{{\rm i}q b \phi} \rangle , \qquad q = \pm 1 , 
\end{equation} 
where the function $R(\theta)$ is given on the interval
$-2\pi + p\pi < {\rm Im}(\theta) < - p\pi$ by the integral
\begin{subequations}
\begin{eqnarray}
R(\theta) & = & {\cal N} \exp \left\{ 8 \int_0^{\infty} \frac{{\rm d}t}{t} 
\frac{\sinh(t) \sinh(p t) \sinh[(1+p)t]}{
\sinh^2(2t)}  \right. \nonumber \\ & &  \left. \times
\sinh^2\left[ t\left( 1 - \frac{{\rm i}\theta}{\pi}
\right) \right] \right\}, \label{5.17a} \\
{\cal N} & = & \exp \left\{ 4 \int_0^{\infty} \frac{{\rm d}t}{t} 
\frac{\sinh(t) \sinh(p t) \sinh[(1+p)t]}{
\sinh^2(2t)} \right\} . \nonumber \\ & & \label{5.17b}
\end{eqnarray}
\end{subequations}
This function satisfies a useful relation
\begin{equation} \label{5.18}
R(\theta) R(\theta\pm {\rm i}\pi) =
\frac{\sinh(\theta)}{\sinh(\theta) \mp {\rm i}\sin(p\pi)} ,
\end{equation}
which, together with $R(-\theta)=S_{11}(\theta )R(\theta )$, 
enables one to extend the definition of $R(\theta)$ to 
arbitrary values of ${\rm Im}(\theta)$. (Here $S_{11}$ denotes the 
$B_1B_1$ scattering matrix, see below).
The evaluation of the one-particle $B_2$ form factor can be based
on a bootstrap procedure \cite{Smirnov}.
Namely, since the $B_2$-breather is a boundstate of the two $B_1$-breathers
(i.e. the $B_1B_1$ scattering matrix has the $B_2$-pole), the
one-particle $B_2$ form factor can be calculated from the two-particle
$B_1$ form factor (\ref{5.16}) as follows \cite{Smirnov}
\begin{equation} \label{5.19}
\Gamma F_1^{(2)}(q) = - {\rm i}\  {\rm res}_{\epsilon = 0}  
F_2^{(11)}\left( \theta + \frac{{\rm i}p\pi + \epsilon }{2} ,
\theta - \frac{{\rm i}p\pi+ \epsilon}{2} ; q \right) , 
\end{equation}
where $\Gamma$ is related to the residue of the $B_2$-pole in the
$B_1B_1$ scattering as follows 
\begin{equation} \label{5.20}
- {\rm i}\  {\rm res}_{\theta={\rm i}p\pi} S_{11}(\theta) = \Gamma^2 .
\end{equation}
Explicitly, one has \cite{Zamolodchikov79}
\begin{equation} \label{5.21}
S_{11}(\theta) = \frac{\sinh(\theta) + {\rm i} \sin(p\pi)}{
\sinh(\theta) - {\rm i} \sin(p\pi)} , \quad
\Gamma=\sqrt{2 \tan(p\pi)} .
\end{equation}
The one-particle $B_2$ form factor thus reads
\begin{equation} \label{5.22}
F_1^{(2)}(q) = - \lambda^2 \left[ \frac{\tan(p\pi)}{2} \right]^{1/2}
\frac{1}{R[{\rm i}\pi(1+p)]} \langle {\rm e}^{{\rm i}qb\phi} \rangle .
\end{equation}

With regard to the bulk neutrality condition
$\langle {\rm e}^{{\rm i}b\phi} \rangle = 
\langle {\rm e}^{-{\rm i}b\phi} \rangle$, under the transformation $q\to -q$ 
the form factor $F_1^{(1)}(q)$ changes the sign, 
$F_1^{(1)}(+1) = - F_1^{(1)}(-1)$, while the form factors 
$F_2^{(11)}(\theta;q)$ (\ref{5.16}) and $F_1^{(2)}(q)$
(\ref{5.22}) are unchanged, $F_2^{(11)}(\theta;+1) = F_2^{(11)}(\theta;-1)$ 
and $F_1^{(2)}(+1) = F_1^{(2)}(-1)$.
Consequently, as is clear from the definitions (\ref{3.18}) and
(\ref{3.19}), within the series representation (\ref{5.13}),
the term with $F_1^{(1)}$ contributes to the charge density $\rho$
while the terms with $F_2^{(11)}$ and $F_1^{(2)}$ contribute to the
particle number density $n$.

\subsection{Asymptotic charge profile}
The asymptotic large-$x$ behavior of the charge density is obtained 
in the form
\begin{eqnarray}
\rho(x) & \displaystyle{\mathop{\sim}_{x\to\infty}} & 2 n
\sqrt{\cos\left( \frac{p\pi}{2} \right)} \left[
1 + \cos\left( \frac{p\pi}{2} \right) - \sin\left( \frac{p\pi}{2}
\right) \right] \nonumber \\
& & \times \exp\left( - \int_0^{p\pi} \frac{{\rm d}t}{2\pi} 
\frac{t}{\sin t} \right) \tanh(p\varphi) \nonumber \\
& & \times \exp\left( - m_1 x \right) , \label{5.23} 
\end{eqnarray}
where $p=\beta/(4-\beta)$ and the mass $m_1$ of the lightest
$B_1$-breather is given by the relations (\ref{2.19}) and (\ref{2.20}).
The formula (\ref{5.23}) applies to the whole stability region
$\beta<2$ which is simultaneously the region of the existence of
the $B_1$- breather in the particle spectrum of the sine-Gordon model.
With respect to the Poisson Eq. (\ref{3.6}), the deviation of
the induced electrostatic potential from its bulk value (\ref{3.7})
behaves at large $x$ as
$\delta \varphi(x) \mathop{\sim}_{x\to\infty} - 2\pi\rho(x)/m_1^2$.
Since the pure exponential asymptotic decay of the DH results
(\ref{4.4}) is recovered, the concept of renormalized charge
is applicable to the present model.
The renormalized image charge $\sigma_{\rm ren}$, defined by
Eq. (\ref{4.18}) with the replacement $\kappa\to m_1$,
is given by
\begin{eqnarray}
\sigma_{\rm ren} & = & - \frac{\kappa}{\pi \beta} 
\sqrt{\frac{p \pi}{2\sin(p\pi/2)}}
\left[ 1 + \cos\left( \frac{p\pi}{2} \right) - \sin\left( \frac{p\pi}{2}
\right) \right] \nonumber \\
& & \times \exp\left( - \int_0^{p\pi} \frac{{\rm d}t}{2\pi} 
\frac{t}{\sin t} \right) \tanh(p\varphi) . \label{5.24}
\end{eqnarray}
It can be readily verified that the $\beta$-expansion (\ref{4.19})
is reproduced by this formula.
In the limits $\beta\to 0$ and $\varphi\to \infty$ such that the
product $\beta\varphi$ is finite, Eq. (\ref{5.24}) reduces to the
result (\ref{4.27}) of the nonlinear PB theory which is therefore
adequate in such regime.
For a given $\beta$, increasing the potential difference
$\varphi$ to infinity, $\sigma_{\rm ren}$ saturates monotonically 
at the finite value 
\begin{eqnarray}
\sigma_{\rm ren}^* & = & - \frac{\kappa}{\pi \beta} 
\sqrt{\frac{p \pi}{2\sin(p\pi/2)}}
\left[ 1 + \cos\left( \frac{p\pi}{2} \right) - \sin\left( \frac{p\pi}{2}
\right) \right] \nonumber \\
& & \times \exp\left( - \int_0^{p\pi} \frac{{\rm d}t}{2\pi} 
\frac{t}{\sin t} \right) , \label{5.25}
\end{eqnarray}
in agreement with the saturation hypothesis.

\subsection{Asymptotic number density profile}
The asymptotic large-$x$ behavior of the particle number density
$n(x)$ is a more complicated topic.
For the density deviation from its bulk value $\delta n(x) = n(x) - n$,
the form-factor asymptotic expansion (\ref{5.13}) gives
\begin{equation} \label{5.26}
\delta n(x) \mathop{\sim}_{x\to\infty} 
\delta n^{(2)}(x) + \delta n^{(11)}(x) ,
\end{equation}
where the $B_2$-breather term reads
\begin{eqnarray}
\delta n^{(2)}(x) & = & - \frac{n}{2} \lambda^2 g_2
\left[ \frac{\tan(p\pi)}{2} \right]^{1/2}
\frac{1}{R[{\rm i}\pi(1+p)]} \nonumber \\
& & \times \exp( - m_2 x) \label{5.27} 
\end{eqnarray}
with $g_2$ given by Eq. (\ref{3.25b}) taken at $\eta = {\rm i}
2\varphi$,
and the $B_1B_1$ integral term reads
\begin{subequations} \label{5.28}
\begin{eqnarray}
\delta n^{(11)}(x) & = & \int_0^{\infty} \frac{{\rm d}\theta}{\pi}
\chi(\theta) {\rm e}^{-2 m_1 x \cosh \theta} , \label{5.28a} \\
\chi(\theta) & = & - \frac{1}{2} n \lambda^2 K^{11}(\theta) 
R(\theta) . \label{5.28b}
\end{eqnarray}
\end{subequations}
The presence of the $B_2$ term is restricted to $p<1/2$ $(\beta<4/3)$, 
the $B_1B_1$ is present in the whole stability region $p<1$ $(\beta<2)$.
As soon as $\beta$ (or, equivalently, $p$) has a strictly nonzero
value, it can be shown from (\ref{5.28}) that  
\begin{equation} \label{5.29}
\delta n^{(11)}(x) \propto \frac{{\rm e}^{-2 m_1 x}}{\sqrt{m_1 x}}
\qquad \mbox{at large $x$,}
\end{equation}
with $2 m_1 > m_2$ for any $\beta>0$.
The $B_1B_1$ term thus becomes subleading in (\ref{5.26}), 
while the $B_2$ term (\ref{5.27}) dominates: 
\begin{equation} \label{5.30}
\delta n(x) \mathop{\sim}_{x\to\infty} 
\delta n^{(2)}(x) , \qquad \beta > 0 .
\end{equation}

As concerns the $\beta\to 0$ limit, considering $R({\rm i}\pi) \to -1$,
$\lambda\to 2\sqrt{p\pi}$, $g_2\to 8\sqrt{2p/\pi^3} [\varphi^2+(\pi^2/16)]$, 
$p\to \beta/4$ and $m_2\to 2\kappa$ in (\ref{5.27}) leads to the expression 
\begin{equation} \label{5.31}
\delta n^{(2)}(x) =  \beta^2 n \left( \varphi^2 + \frac{\pi^2}{16}
\right) \exp ( -2 \kappa x )  + O(\beta^3) ,
\end{equation}
which is twice larger than the expected result (\ref{4.22}) of the
systematic $\beta$-expansion.
This means that the formula (\ref{5.30}) does not reflect 
adequately the expansion of the asymptotic $\delta n(x)$
around the $\beta=0$ point.
The reason for this inconsistency consists in the fact that, 
in the limit $p\to 0$, besides the important equality 
of inverse correlation lengths $m_2 = 2 m_1 = 2\kappa$,
the value of $\chi(\theta)$ as well as of all its derivatives 
go to infinity at $\theta=0$.
As a consequence, also the $B_1B_1$ term (\ref{5.28a}), which is
subleading for a strictly nonzero $\beta$, contributes to the leading
order of the pure-exponential large-$x$ behavior [see the relation
(\ref{5.35}) below].
It is important to add that if we would be able to perform
all $\beta$-orders of the large-$x$ decay of $\delta^{(11)}(x)$, 
we should arrive at the exact asymptotic behavior (\ref{5.29}),
valid for $\beta>0$, which is no longer purely exponential. 
This mathematical technicality was first observed in the study of 
finite-size effects for the (1+1)-dimensional sine-Gordon
theory defined on a strip with Dirichlet-type boundary conditions
\cite{Bajnok05}; the next analysis follows a regularization
procedure presented in Ref. \onlinecite{Bajnok05}.
The dangerous singularity of $\chi(\theta)$, at $\theta=0$ in
the limit $p\to 0$, can be isolated from $\chi(\theta)$ in
the following way 
\begin{equation} \label{5.32}
\chi(\theta) = \frac{\cosh \theta +\cos(p\pi/2)}{
\cosh \theta -\cos(p\pi/2)} \chi_0(\theta) ,
\end{equation}
where $\chi_0(\theta)$ is a regular function of $\theta$ around
$\theta=0$:
\begin{equation} \label{5.33}
\chi_0(\theta) = \chi_0(0) + \chi_0''(0) \frac{\theta^2}{2} + \cdots .
\end{equation}
Here, $\chi_0(0)$ corresponds to the classical treatment,
$\chi_0''(0)$ to the first quantum correction, etc.
Within the regularized form (\ref{5.32}) complemented by the regular
expansion (\ref{5.33}), the evaluation of the $B_1B_1$ integral (\ref{5.28a}) 
can be carried out in close analogy with Ref. \onlinecite{Bajnok05}.
In particular, for very large $x$ and small $p$, one uses the results of 
\onlinecite{Bajnok05}:
\begin{subequations}
\begin{eqnarray}
\int_0^{\infty} \frac{{\rm d}\theta}{\pi}
\frac{\cosh \theta + \cos(p\pi/2)}{\cosh \theta - \cos(p\pi/2)}
{\rm e}^{-2 m x \cosh \theta}& & \nonumber \\  
& &\hspace{-5.5cm}\displaystyle{\mathop{\sim}_{x\to\infty}}\quad  
\frac{\cos(p\pi/2)}{\sin(p\pi/4)} {\rm e}^{-2 m x} . \label{5.34a} 
\end{eqnarray}
Similarly, one can derive that 
\begin{eqnarray}
\int_0^{\infty} \frac{{\rm d}\theta}{\pi}
\frac{\cosh \theta + \cos(p\pi/2)}{\cosh \theta - \cos(p\pi/2)}
\frac{\theta^2}{2} {\rm e}^{-2 m x \cosh \theta} & &\nonumber \\ 
& &\hspace{-5.5cm}\displaystyle{\mathop{\sim}_{x\to\infty}} \quad  
2 \sin \left( \frac{p\pi}{4} \right) 
\cos \left( \frac{p\pi}{2} \right) {\rm e}^{-2 m x} . \label{5.34b}
\end{eqnarray}
\end{subequations}
Regarding the explicit forms of $\chi_0(0)$ and $\chi_0''(0)$, 
the result for the $B_1B_1$ integral is such that
\begin{equation} \label{5.35}
\lim_{p\to 0} \frac{\delta n^{(11)}(x)}{\delta n^{(2)}(x)}
= - \frac{1}{2} .
\end{equation}
In view of relation (\ref{5.31}), the sum in Eq. (\ref{5.26}) 
thus reproduces the needed result (\ref{4.22}).
To conclude, the small $\beta$-expansion around the $\beta=0$ point
of the asymptotic number density profile involves artificial 
contributions from the $B_1B_1$ integral term, this term being totally 
absent at a strictly nonzero (possibly very small) value of $\beta$.
In other words, the small-$\beta$ expansion of the asymptotic
density profile, when truncated at some finite $\beta$-order, 
does not reflect adequately the asymptotic density profile at
a nonzero value of $\beta$.
From this point of view, the DH results for number density profiles have to
be taken with caution also for other boundary Coulomb systems.

\renewcommand{\theequation}{6.\arabic{equation}}
\setcounter{equation}{0}

\section{The free-fermion $\beta=2$ point}
For the sake of completeness, we summarize in view of the subjects 
of present interest the exact results for the 2D electrical double
layer at the free-fermion $\beta=2$ point of the Thirring
representation of the Coulomb gas \cite{Cornu2}.
The exact solution for the species density profiles at an
arbitrary distance $x>0$ from the ideal-conductor wall reads
\begin{eqnarray} 
n_{\pm}(x) - n_{\pm} & = & \frac{m}{2\pi} \int_0^{\infty} {\rm d}l
\left[ - \frac{m}{\kappa_l} + \frac{\kappa_l 
\exp(\pm\beta\varphi)+m}{m \cosh (\beta\varphi)+\kappa_l} \right]
\nonumber \\ & & \qquad \times \exp(-2\kappa_l x) , \label{6.1}
\end{eqnarray} 
where $m=2\pi z$ is the rescaled fugacity (equal to the soliton
mass M, see formula (\ref{2.17}) taken at $b^2=1/2$ and $p=1$),
$\kappa_l = (m^2+l^2)^{1/2}$ and $n_{\pm}=n/2$ are the bulk densities
regularized by considering a hard-core repulsion around each particle.
The short-distance limit of Eq. (\ref{6.1}),
\begin{equation} \label{6.2}
n_{\pm}(x) \sim \frac{z_{\pm}}{2 x} \quad \mbox{as $x\to 0$} ,
\end{equation}
is of the expected form (\ref{3.12}).
The charge density, calculated from Eq. (\ref{6.1}), implies via
the Poisson equation the following deviation of the electrostatic
potential from its bulk value:
\begin{equation} \label{6.3}
\delta\varphi(x) =  - \int_0^{\infty} {\rm d} l
\frac{1}{2\kappa_l} \frac{m \sinh(2\varphi)}{m 
\cosh (2\varphi) + \kappa_l} \exp(-2\kappa_l x) . 
\end{equation}

At asymptotically large distances from the wall,
\begin{equation} \label{6.4}
\delta\varphi(x) \mathop{\sim}_{x\to\infty} - \frac{1}{4}
\tanh \varphi \left( \frac{\pi}{m x} \right)^{1/2}
\exp(-2 m x) .
\end{equation}
This asymptotic behavior differs fundamentally from the purely
exponential DH prediction (\ref{4.4a}).
The concept of renormalized charge is therefore not applicable
at the free fermion point.
The reason for the fundamental difference is obvious.
The lightest $B_1$-breather disappears from the particle spectrum
of the sine-Gordon model just at the free-fermion point $\beta=2$,
and the asymptotic behavior of $\delta\varphi(x)$ starts to be
governed by the soliton-antisoliton pair.
Since $m_1\to 2M$ $(\equiv 2m)$ as $\beta\to 2$, the particle mass
in the exponential decay on the rhs of Eq. (\ref{6.4}) is a continuous
function of $\beta$ at $\beta=2$.
On the other hand, the position-dependent prefactor $(mx)^{-1/2}$ 
in the formula (\ref{6.4}), determined by the form-factor of 
the soliton-antisoliton pair, has no ``continuous'' analogue in 
the leading asymptotic behavior of $\delta\varphi(x)$ inside 
the stability region $\beta<2$.
The basic qualitative features of the results at the $\beta=2$ point
are expected to be present also for $\beta>2$, up to the 
Kosterlitz-Thouless point $\beta=4$ where the 2D sine-Gordon theory
ceases to be massive.

In the limit $\varphi\to\infty$, $\delta\varphi(x)$ of Eq. (\ref{6.3})
saturates at
\begin{equation} \label{6.5}
\delta\varphi^*(x) = - \frac{1}{2} K_0(2 m x) .
\end{equation}
This function is finite in the whole electrolyte region $x>0$,
which confirms the validity of the hypothesis of the electric
potential saturation \cite{Tellez}.

\renewcommand{\theequation}{7.\arabic{equation}}
\setcounter{equation}{0}

\section{Conclusion}
The main aim of this paper was to test basic concepts used in the
theory of highly asymmetric Coulomb fluids on the exact solution
of a 2D electrical double layer.
This model is mappable onto the 2D semi-infinite sine-Gordon
field theory with Dirichlet boundary conditions which do not
break the integrability property of the bulk sine-Gordon model
with the known particle spectrum.
At large distances from model's interface, the induced
electric potential has the pure exponential decay for small 
enough inverse temperatures (couplings) $\beta<2$, including the DH 
$\beta\to 0$ limit.
This fact confirms the adequacy of the concept of renormalized
charge for weak couplings.
In the extreme case of an infinite potential difference between
model's interface and the bulk interior of the electrolyte, 
the renormalized charge saturates at a finite value which is 
in agreement with the saturation hypothesis.

As concerns the applicability of the present results in the
context of ``real'' 3D Coulomb fluids, we expect the reduced 
dimensionality to be unimportant because, as soon as one uses 
the 2D harmonic (i.e., logarithmic) potential, basic screening sum
rules for the charge-density averages are maintained. 
Another ``imperfection'' of our model is the absence of a hard core
around each charged particle: in 3D, the hard-core repulsion is
inevitable to avoid the thermodynamic collapse at any $\beta>0$.
Since the charge sum rules are not affected by a short-ranged
regularization of the Coulomb potential \cite{Martin}, also
this model's simplification seems to be irrelevant. 

\begin{acknowledgments}
The support by an exchange program between Hungarian and Slovak
Academies of Sciences is acknowledged. 
L. {\v S}. was supported by Grant VEGA 2/3107/24 and Z. B. by OTKA T037674,
T034299, T043582 and the EC network EUCLID HPRN-CT-2002-00325. 
\end{acknowledgments}

\end{document}